\documentclass{sig-alternate}
\toappearbox{In submission. Last modified 3 Jan 2013.}  %

\usepackage{amssymb,amsmath}
\usepackage{bbm}

\usepackage{algorithm, algorithmicx}
\usepackage{algpseudocode}

\usepackage{graphicx}
\usepackage{caption}
\usepackage{subfig}
\usepackage{float}

\usepackage{multirow}
\usepackage{booktabs}

\usepackage{url}

\usepackage{enumitem}

\usepackage[numbers]{natbib}
\newcommand{\bp}{\mathbf{p}}
\newcommand{\R}{\mathbb{R}}
\newcommand{\comment}[1]{}

\begin{document}

\title{Word Storms: Multiples of Word Clouds for \\ Visual Comparison of Documents}

\numberofauthors{2} %
\author{
\alignauthor
Quim Castella\\ %
	\affaddr{School of Informatics}\\
       \affaddr{University of Edinburgh}\\
       \email{quim.castella@gmail.com}
\alignauthor
Charles Sutton\\
	\affaddr{School of Informatics}\\
       \affaddr{University of Edinburgh}\\
       \email{csutton@inf.ed.ac.uk}
}

\maketitle
\begin{abstract}
Word clouds are a popular tool for visualizing documents, but they are not
a good tool for comparing documents, because identical words are not
presented consistently across different clouds.
We introduce the concept of \emph{word storms}, 
a visualization tool for analysing corpora of documents. 
A word storm is a group of word clouds, in which each cloud
represents a single document, juxtaposed to allow the viewer
 to compare and contrast the documents.
We present a novel algorithm that creates a coordinated word storm, 
in which words that appear in multiple documents are placed in the same location, 
using the same color and orientation, in all of the corresponding clouds. 
In this way, similar documents are represented by similar-looking word clouds,
making them easier to compare and contrast visually.
We evaluate the algorithm in two ways: first, an automatic evaluation
based on document classification; and second, a user study.
The results confirm that unlike standard word clouds, 
a coordinated word storm better allows for visual comparison of documents.
\end{abstract}


\section{Introduction}

Because of the vast number of text documents on the Web,
there is a demand for ways to allow  people to scan large numbers
of documents quickly.  A natural approach is visualization, 
under the hope that visually scanning a picture may be easier for people than reading text.
One of the most popular visualization methods for text 
documents are \emph{word clouds}. A word cloud is a graphical presentation
of a document, usually generated by plotting the document's most common
words in two dimensional space,  with the word's frequency indicated by its font size.
Word clouds have the advantages that they are easy for
naive users to interpret and that they can be 
aesthetically surprising and pleasing.
One of the most popular cloud generators, Wordle,
has generated over 1.4 million clouds that have been publicly
posted \cite{feinberg10beautiful}.

Despite their popularity for visualizing single documents,
word clouds are not useful for navigating groups of documents,
such as blogs or Web sites. The key problem is that word clouds
are difficult to compare visually. For example, say that we want to compare two documents,
so we build a word cloud separately for each document.
Even if the two documents are topically similar, the resulting clouds can be 
very different visually, because the shared words between the documents 
are usually scrambled, appearing in different locations in each of the two clouds.
The effect is that it is difficult to determine which words are shared between the documents.

In this paper, we introduce the concept of \emph{word storms} to afford
visual comparison of groups of documents.
Just as a storm is a group of clouds, a word storm is a group of word clouds.
Each cloud in the storm represents a subset of the corpus. For example,
a storm might contain one cloud per document, or alternatively one cloud to represent all the documents
written in each year, or one cloud to represent each track of an academic conference,
etc. Effective storms make it easy to compare and contrast documents visually.
We propose several principles behind effective storms, the most important
of which is that
\emph{similar documents should be represented by visually  similar clouds.}
To achieve this, algorithms for generating storms must perform layout
of the clouds in a coordinated manner.

We present a novel algorithm for generating coordinated word storms.
Its goal is to generate a set of visually appealing clouds, under the constraint
that if the same word appears in more than one cloud in the storm, it appears in a similar 
location. Interestingly, this also allows a user to see when a word is \emph{not} in a cloud: simply find the desired word in one cloud and check the corresponding
locations in all the other clouds.
At a technical level, our
algorithm combines the greedy randomized layout strategy of Wordle,
which generates aesthetically pleasing layouts, with an optimization-based
approach to maintain coordination between the clouds.  The objective function in the optimization
measures the amount of coordination in the storm and
is inspired by the theory of multidimensional scaling.

We apply this algorithm on a variety of text corpora, including academic papers
and research grant proposals.  We evaluate the algorithm in two ways.
First, we present a novel automatic evaluation method for word storms based on how well the
clouds, represented as vectors of pixels, serve as features for document classification.
The automatic evaluation allows us to rapidly compare different layout algorithms.
However, this evaluation is not specific to word clouds and may be of independent interest.
Second, we present a user study in which users are asked to
examine and compare the clouds in storm.
Both experiments demonstrate that a coordinated word storm is dramatically better
than independent word clouds at allowing users to visually compare
and contrast documents.

%

\section{Design Principles}

In this section we introduce the concept of a word storm,
describe different types of storms, and present
design principles for effective storms.
 
A word storm is a group of word clouds constructed for the purpose of
visualizing a corpus of documents.
In the simplest type of storm,
each cloud represents a single document by creating a summary of its content; hence, by looking at the clouds a user can form a quick
impression of the corpus's content and analyse the relations among the different documents.
 
We build on word clouds in our work because they are a popular way of visualising single documents.
They are very easy to understand and they have been widely used to create appealing figures.
By building a storm based on word clouds, we create an accessible tool that can be understood easily and used without requiring a background in statistics.
The aim of a word storm is to extend the capabilities of a word cloud: instead of visualizing just one document, it is used to
visualize an entire corpus.

There are two high level design motivations behind the concept of word storms.
The first design motivation is to \emph{visualize high-dimensional data in a high-dimensional space}.
Many classical visualization techniques are based on dimensionality reduction,
i.e., mapping high-dimensional data into a low dimensional space.
Word storms take an alternative strategy, of mapping high dimensional
data into a different high dimensional space, but one which is
tailored for human visual processing.
This a similar strategy to approaches like Chernoff faces \cite{chernoff73faces}.
The second design motivation is the \emph{principle of small multiples}
\cite{Tufte2001,Tufte1997}, in which similar visualizations are presented
together in a table so that the eye is drawn to the similarities and
differences between them.  A word storm is a small multiple of word clouds.
This motivation strongly influences the design of effective clouds,
as described in Section~\ref{sec:goodstorm}.

\begin{figure*}[tb!]
\centering
\subfloat[]{\fbox{\includegraphics[width=0.22\textwidth]{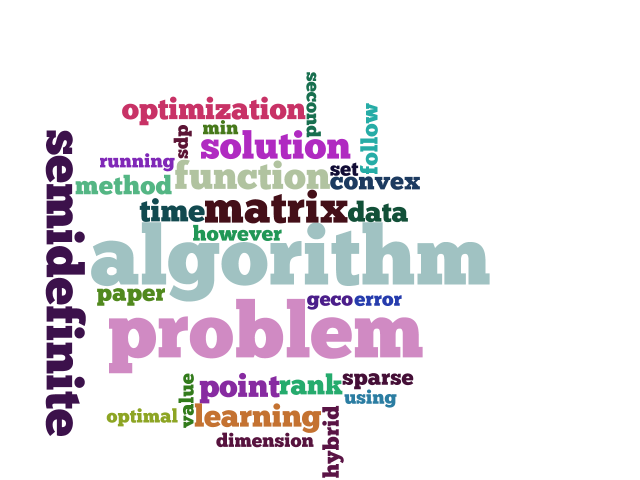}}}\hspace{1pt}
\subfloat[]{\fbox{\includegraphics[width=0.22\textwidth]{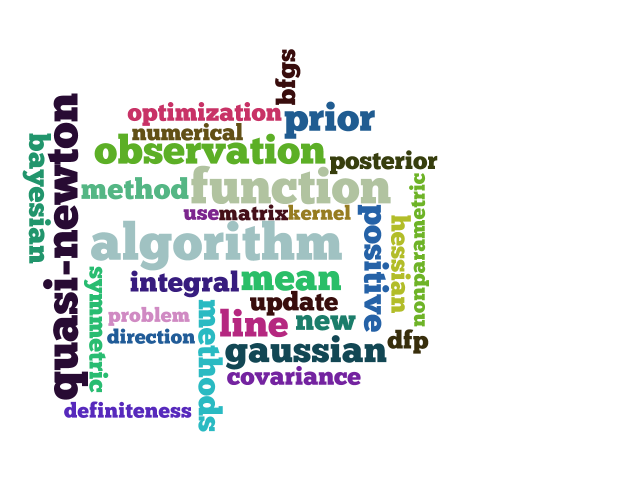}}}\hspace{1pt}
\subfloat[]{\fbox{\includegraphics[width=0.22\textwidth]{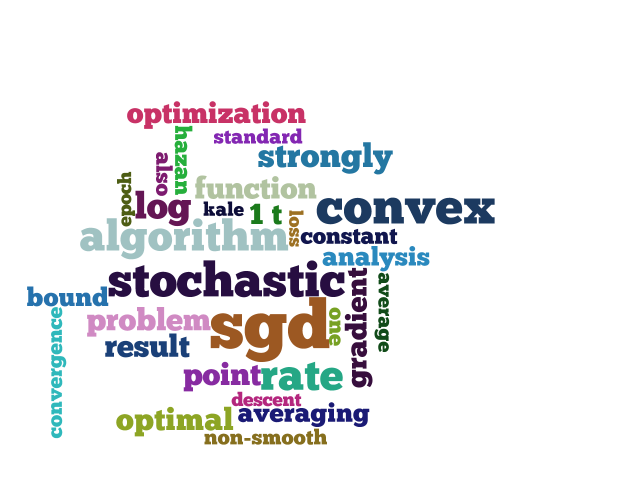}}}\hspace{1pt}
\subfloat[]{\fbox{\includegraphics[width=0.22\textwidth]{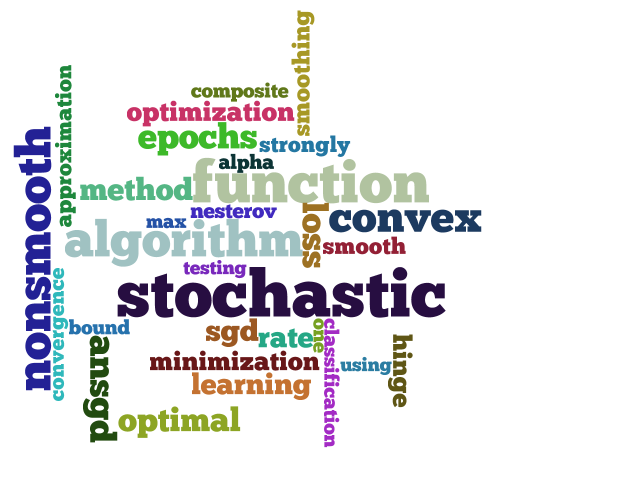}}}\vspace{1pt}
\subfloat[]{\fbox{\includegraphics[width=0.22\textwidth]{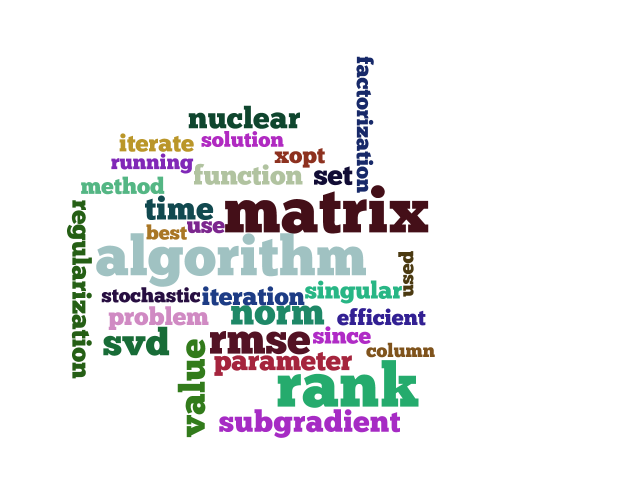}}}\hspace{1pt}
\subfloat[]{\fbox{\includegraphics[width=0.22\textwidth]{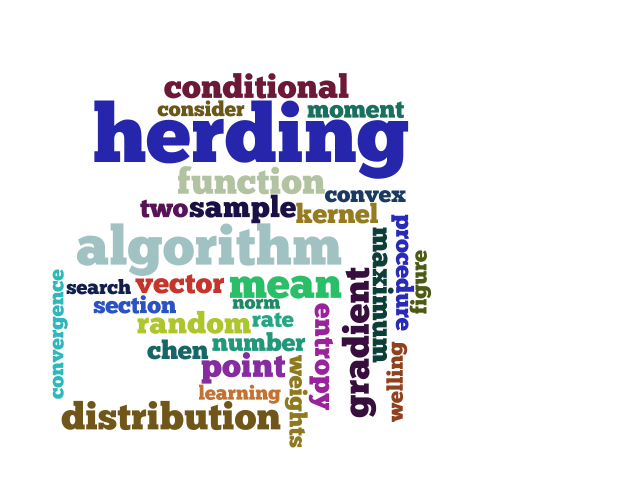}}}\hspace{1pt}
\subfloat[]{\fbox{\includegraphics[width=0.22\textwidth]{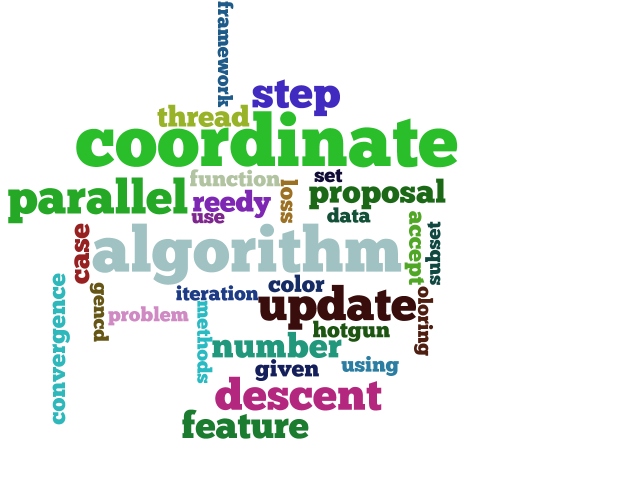}}}\hspace{1pt}
\subfloat[]{\fbox{\includegraphics[width=0.22\textwidth]{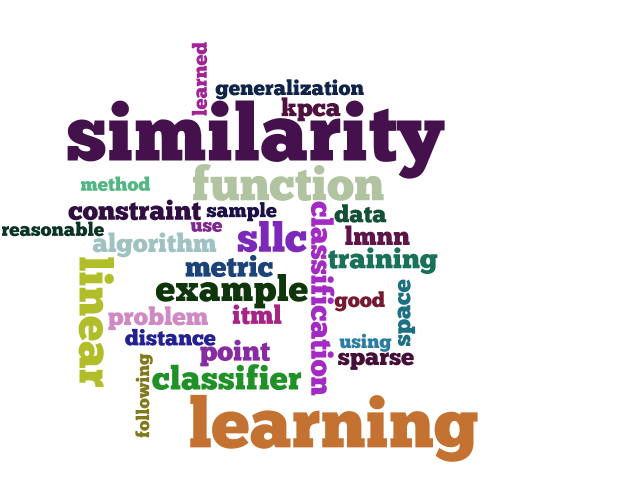}}}
\caption{We represented the papers of the ICML 2012 conference. These 8 clouds represent the papers in the Opitmization Algorithms track.}
\label{fig:icml}
\end{figure*}

\subsection{Types of Storms}

Different types of storms can be constructed for different data analysis tasks.
In general, the individual clouds in a storm can represent a group of
documents rather than a single document.
 For example, a cloud could represent all the documents
written in a particular month, or that appear on a particular section
of a web site. It would be typical to do this by simply merging
all of the documents in each group, and then generating the storm with
one cloud per merged document.
This makes the storm a flexible tool that can be used for
different types of analysis, and it is possible
to create different storms from the same corpus and 
obtain different insights on it.
Here are some example scenarios:

\begin{enumerate}[leftmargin=*]
 \item \textbf{Comparing Individual Documents}. 
 If the goal is to compare and contrast individual documents
 in a corpus, then we can build in a storm in which each word cloud 
 represents a single document. 

 \item \textbf{Temporal Evolution of Documents}.
If we have a set of documents that have been written over a long period,
such as news articles, blog posts, or scientific documents,
we may want to analyze how trends in the documents have changed over time.
This is achieved using a word storm in which each cloud represents a time period,
e.g., one cloud per week or per month. By looking at the clouds sequentially, the user can see the appearance and disappearance of words and 
how their importance changes over time.

 \item \textbf{Hierarchies of Documents}.
If the corpus is arranged in a hierarchy of categories, we can create a storm which contains one cloud for each of the categories and subcategories.
This allows for hierarchical interaction, in which for every category of the topic
hierarchy, we have a storm that contains one cloud for each subcategory.
For instance, this structure can be useful in a corpus of scientific papers. At the top level, we would first have a storm that contains one cloud for each scientific
field (e.g., chemistry, physics, engineering), then for each field, we also have a separate storm that includes one cloud for each subfield (such as organic chemistry, inorganic chemistry) and so on until arriving at the articles.
An example of this type of storm is shown in Figures~\ref{fig:epsrc1} and~\ref{fig:epsrc2}.

\begin{figure*}[p]
\centering
\subfloat[Chemistry]	{\fbox{\includegraphics[width=0.31\textwidth]{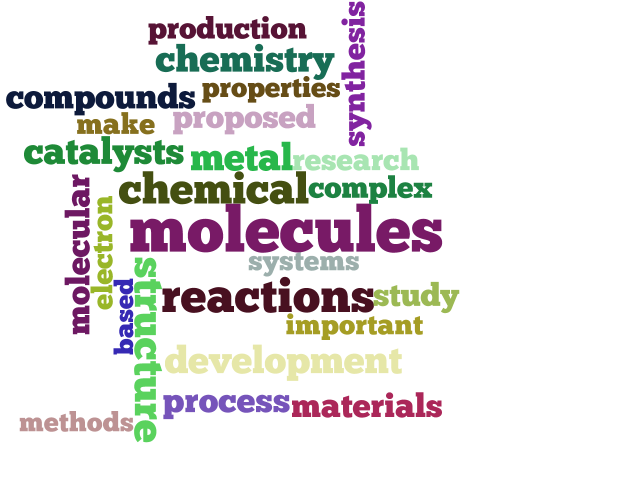}}}\hspace{1pt}
\subfloat[Engineering]	{\fbox{\includegraphics[width=0.31\textwidth]{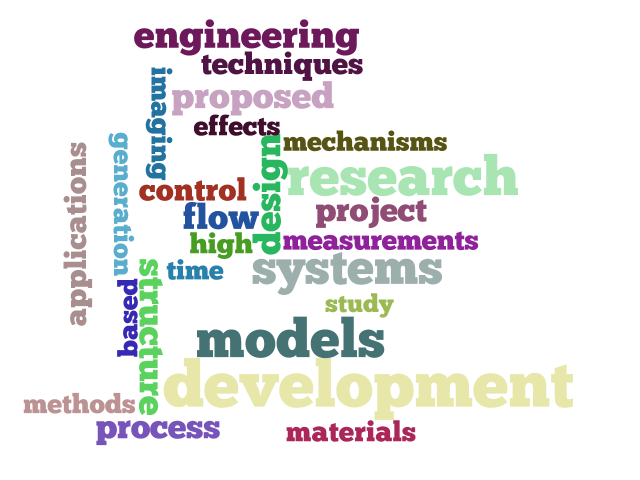}}}\hspace{1pt}
\subfloat[Information Communication and Technology]		{\fbox{\includegraphics[width=0.31\textwidth]{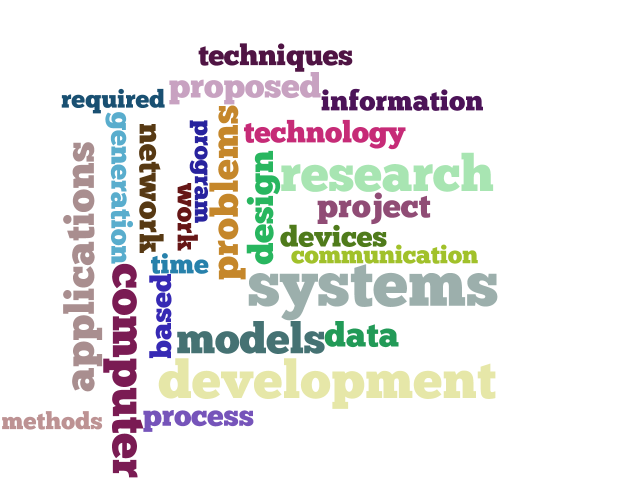}}}\vspace{1pt}
\subfloat[Physical Sciences]	{\fbox{\includegraphics[width=0.31\textwidth]{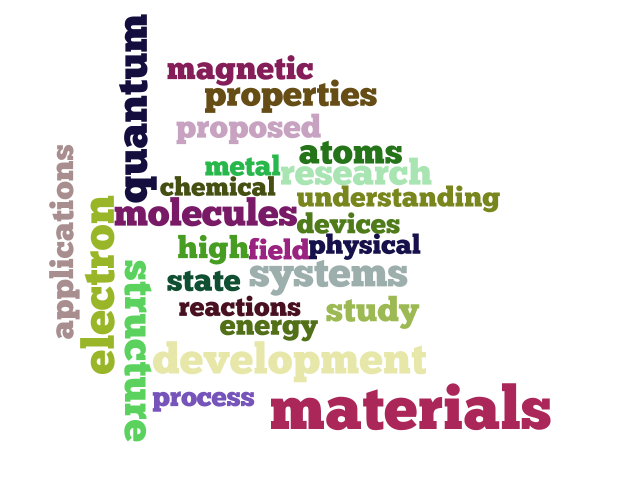}}}\hspace{1pt}
\subfloat[Complexity]	{\fbox{\includegraphics[width=0.31\textwidth]{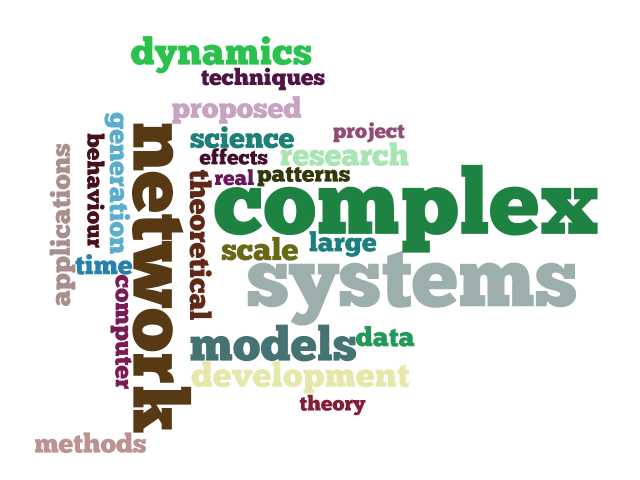}}}\hspace{1pt}
\subfloat[Mathematical Sciences]{\fbox{\includegraphics[width=0.31\textwidth]{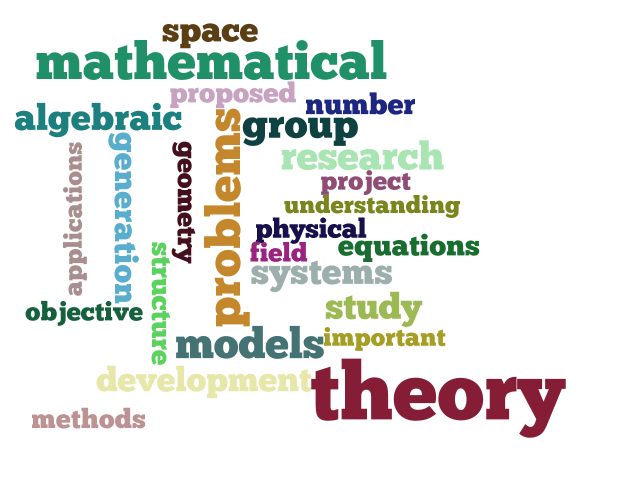}}}
\caption{These clouds represent 6 EPSRC Scientific Programmes. Each of the programmes is obtained by concatenating all its grants abstracts.}\label{fig:epsrc1}
\end{figure*}
\begin{figure*}[p]
\centering
\subfloat[]{\fbox{\includegraphics[width=0.31\textwidth]{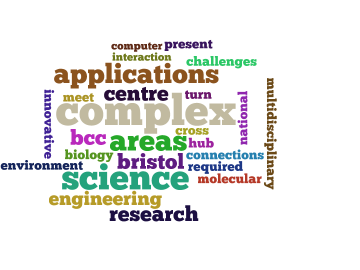}}}\hspace{1pt}
\subfloat[]{\fbox{\includegraphics[width=0.31\textwidth]{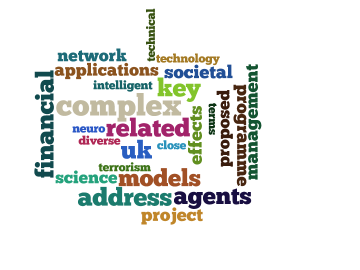}}}\hspace{1pt}
\subfloat[]{\fbox{\includegraphics[width=0.31\textwidth]{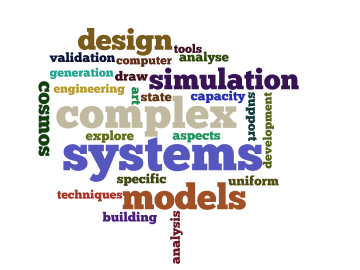}}}\vspace{1pt}
\subfloat[]{\fbox{\includegraphics[width=0.31\textwidth]{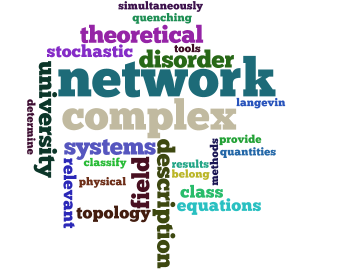}}}\hspace{1pt}
\subfloat[]{\fbox{\includegraphics[width=0.31\textwidth]{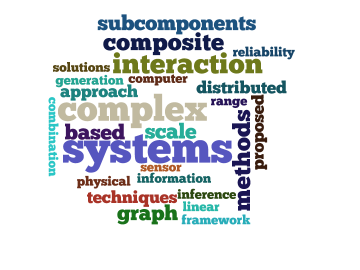}}}\hspace{1pt}
\subfloat[]{\fbox{\includegraphics[width=0.31\textwidth]{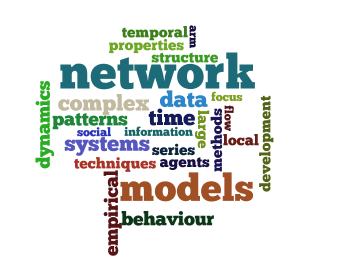}}}
\caption{A word storm containing six randomly sampled grants from the Complexity Programme (which was Cloud (e) in Figure \ref{fig:epsrc1}). 
The word ``complex'', that only appeared in one cloud in Figure \ref{fig:epsrc1}, 
appears in all clouds in this Figure.
As this word conveys more information in Figure \ref{fig:epsrc1} than in here, 
here it is colored more transparent.}\label{fig:epsrc2}
\end{figure*}

\end{enumerate}
To keep the explanations simple, when describing the algorithms later on,
we will assume that each cloud in the storm represents a single document,
with the understanding that the ``document'' in this context may have
been created by concatenating a group of documents, as in the storms
of type  2 and 3 above.

\subsection{Levels of Analysis of Storms}

A single word storm allows the user to analyse the corpus at a variety
of different levels, depending on what type of information is of most interest,
such as:
\begin{enumerate}[leftmargin=*] 
 \item \textbf{Overall Impression}. By scanning the largest terms 
 across all the clouds,
 the user can form a quick impression of the topics of whole corpus.

 \item \textbf{Comparison of Documents}. As the storm displays the clouds together, the user can easily compare them and
look for similarities and differences among the clouds. For example, the user
can look for words that are much more common in one document than another.
Also the user can compare whether two clouds have similar shape,
to gauge the overall similarity of the corresponding documents.

 \item \textbf{Analysis of Single Documents}. Finally, the clouds in the storm have
meaning in themselves. Just as with a single word cloud, the user can analyze an
individual cloud to get an impression of a single document.
\end{enumerate}

\subsection{Principles of Effective Word Storms}
\label{sec:goodstorm}

Because they support additional types of analysis, 
principles for effective word storms are different than
those for individual clouds.
This section describes some desirable properties of effective word storms.

First of all, \emph{each cloud should be a good representation of its document}. 
That is, each cloud ought to emphasize the most important words so that the information 
that it transmits is faithful to its content.  Each cloud in a storm
should be an effective visualization in its own right.

Further principles follow from the fact that the clouds should also be built taking into account the roles they will play in the complete storm.
In particular, \emph{clouds should be designed so that they are
effective as small multiples} \citep{Tufte1997,Tufte2001},
that is, they should be easy to compare and contrast.
This has several implications.
First, clouds should be similar so that they look like multiples of the same thing, making the storm a whole unit. Because the same structure is 
maintained across the different clouds,
they are easier to compare, so that the viewer's
attention is focused on the differences among them.
A related implication is that {the clouds ought to be small enough that 
viewers can analyze multiple clouds} at the same time
without undue effort.

The way the clouds are arranged and organised on the canvas can
also play an important role, because clouds are probably
 more easily compared to their neighbours than to the more distant
clouds.  This suggests a principle that \emph{clouds in a storm should be 
arranged to facilitate the most important comparisons}.
In the current paper, we take a simple approach to this issue, simply arranging
the clouds in a grid, but in future work 
it might be a good option to place similar clouds closer together so that they can be 
more easily compared.

A final, and perhaps the most important, principle is one that we will call
\emph{the coordination of similarity} principle. In an effective storm,
visual comparisons between clouds should reflect the underlying relationships
between documents, so that \emph{similar documents should have similar clouds, and dissimilar documents should have visually distinct clouds.}
This principle has particularly strong implications.
For instance, to follow this principle, words should appear in a similar font
and similar colours when they appear in multiple clouds.
More ambitiously, words should also have approximately the same
\emph{position} when the same position across all clouds.

Following the coordination of similarity principle can
significantly enhance the usefulness of the storm.
For example, a common operation when comparing word clouds is to finding and comparing words between the clouds, e.g., once a word is spotted in a cloud, checking
if it also appears in other clouds.
By displaying shared words in the same color and position across clouds, 
it is much easier for a viewer
to determine which words are shared across clouds,
and which words appear in one cloud \emph{but not} in another.
Furthermore, making common words look the same tends to cause the overall
shape of the clouds of similar documents to appear visually similar,
allowing the viewer to assess the degree of similarity of two documents
without needing to fully scan the clouds.

Following these principles presents a challenge for algorithms that build
word storms. 
Existing algorithms for building single word clouds do not take into account
relationships between multiple clouds in a storm. In the next sections
we propose new algorithms for building effective storms.

\section{Creating a Single Cloud}
\label{sec:single}

In this section, we describe the layout algorithm for single clouds that
we will extend when we present our new algorithm for word storms.  
The method is based closely on that of Wordle \citep{feinberg10beautiful}, because
it tends to produce aesthetically pleasing clouds.
Formally, we define a word cloud as a set of words $W = \{w_1,\dots,w_M\}$,
where each word $w \in W$ is assigned a position $p_{w} = (x_{w}, y_{w})$
and visual attributes that include its font size $s_{w}$, color $c_{w}$ and orientation $o_{w}$ (horizontal or vertical).

To select the words in a cloud, we choose the top $M$ words from the document by
term frequency, after removing stop words.  A more general measure of the weight
of each term, such as \emph{tf}*\emph{idf}, could be used instead; for this reason
we use term \emph{weight} to refer to whatever measure we have selected for the
importance of each term.
The font size is set proportional to the term's frequency, and the color and 
orientation are selected randomly.

Choosing the word positions is more complex, because the words must not
overlap on the canvas. We use the layout algorithm from Wordle
\citep{feinberg10beautiful}, which we will refer to as the Spiral Algorithm.

This algorithm is greedy and incremental; it sets the location of one word at a time
in order of weight.
In other words, at the beginning of the $i$-th step, the algorithm has 
generated a partial word cloud containing the $i-1$ words of largest weight.
To add a word $w$ to the cloud, the algorithm places it at an initial desired position $p_{w}$ (e.g., chosen randomly). If at that position, $w$ does not
intersect any previous words and is entirely within the frame, we go on to the next word.
Otherwise, $w$ is moved one step outwards along a spiral path.
The algorithm keeps moving the word over the spiral until it finds a valid position, that is,
it does not overlap and it is inside the frame. Then, it moves on to the next word.
This algorithm is shown in Algorithm~\ref{alg:0}.

We set the word desired positions randomly by sampling a 2D Gaussian distribution whose mean is at the center of the word cloud frame.
The variance is adjusted depending on the width and the height of the word cloud frame. 
If the desired position is sampled outside the frame or it intersects with it, it is resampled until it is inside.

Note that the algorithm assumes that the size of the frame is given.
To choose the size of the frame, 
we estimate the necessary width and the height to fit $M$ words.
This choice will affect the compactness of the resulting word cloud.
If the frame is too big, the words will find valid locations quickly but the resulting
cloud will contain a lot of white space.
If it is frame is too small, it will be more difficult or impossible to
fit all the words.
A maximum number of iterations is set to prevent words from looping forever. 
If one word reaches the maximum number of iterations, we assume that the word cannot fit in the current configuration. 
In that case, the algorithm is restarted with a larger frame.

\begin{algorithm}[h]
\caption{Spiral Algorithm}\label{alg:0}
\begin{algorithmic}[1]
\Require Words $W$, optionally positions $\bp = \{ p_w \}_{w \in W}$  %
\Ensure Final positions $\bp = \{ p_w \}_{w \in W}$
\ForAll{words $w \in \{w_1,\dots, w_M\}$}
	\State \mbox{if initial position $p_w$ unsupplied, sample from Gaussian}
	\State count $\gets$ 0
	\While{ $p_{w}$ not valid $\wedge$ count $<$ Max Iteration }
		\State Move $p_{w}$ one step along a Spiral path
		\State count $\gets$ count + 1
	\EndWhile
	\If { $p_{w}$ not valid}
		\State Restart with a larger Frame
	\EndIf
\EndFor
\end{algorithmic}
\end{algorithm}

In order to decide if two words intersect, we check them at the glyph level, instead
of only considering a bounding box around the word.
This ensures a more compact result.
However, checking the intersection of two glyphs can be expensive, 
so instead we use a tree of rectangular bounding boxes that closely
follows the shape of the glyph, as in
\cite{feinberg10beautiful}. We use the implementation of the
approach in the open source library WordCram.\footnote{\url{http://wordcram.org}}

\section{Creating a Storm}
\label{ch:creatingstrom}

In this section, we present novel algorithms to build a storm.
The simplest method would of course be to simply run the
single-cloud algorithm of Section~\ref{sec:single} independently for
each document, but the resulting storms would typically
violate the principle of coordination of similarity (Section~\ref{sec:goodstorm})
because words will tend to have different colors, orientations, and layouts 
even when they are shared between documents.
Instead, our algorithms will coordinate the layout of different clouds,
so that when words appear in more than one cloud, they have the same
color, orientation, and position.  In this way, if the viewer finds a word
in one of the clouds, it is easy to check if it appears in any other clouds.

We represent each document as a vector $u_i$, where  $u_{iw}$ is the count of word $w$ in document $i$.
A word cloud $v_i$ is a tuple $v_i = (W_i, \{ p_{iw} \}, \{ c_{iw} \},
\{ s_{iw} \})$, where $W_i$ is the set of words that are to be displayed in cloud $i$, and for any word $w \in W_i$, we define $p_{iw} = (x_{iw}, y_{iw})$ as the position of $w$ in the cloud $v_i$,
$c_{iw}$ the color, and $s_{iw}$ the font size.
We write $\bp_i = \{ p_{iw} \,|\, w \in W_i \}$ for the 
set of all word locations in $v_i$.

Our algorithms will focus
on coordinating word locations and attributes of words that are shared
in multiple clouds in a storm. However, it is also possible
to select the words that are displayed in each cloud
in a coordinated way that considers the entire
corpus.
For example, instead of selecting words by their frequency in the current document, 
we could use global measures, such as $tf*idf$, 
that could emphasize the differences among clouds.
We tried a few preliminary experiments with this but subjectively preferred storms
produced using $tf$.

\subsection{Coordinated Attribute Selection}

A simple way to improve the coordination of the clouds in a storm
is to ensure that words that appear in more than one clouds are displayed
with the same color and orientation across clouds.
We can go a bit farther than this, however,
by encoding information in the words' color and orientation. 
In our case, we decided to use color as an additional way of encoding
the relevance of a term in the document.
Rather than encoding this information in the hue, which would required
a model of color saliency, instead we control the color transparency.
We choose the alpha channel of the color to correspond to the inverse document frequency $idf$ of the word in the corpus.
In this way, words that appear in a small number of documents will have opaque colors, while words that occur in many documents will be more transparent.
In this way the color choice emphasizes differences among the documents, 
by making more informative words more noticeable.

\subsection{Coordinated Layout: Iterative Algorithm}

Coordinating the positions of shared words is much more difficult than
coordinating the visual attributes. In this section we present the first
of three algorithms for coordination word positions.
In the same manner that we have set the color and the orientation, 
we want to set the position $p_{wi} = p_{wj}$ $\forall v_i, v_j \in \mathcal{V}_w$, 
where $\mathcal{V}_w$ is the set of clouds that contain word $w$.
The task is more challenging because it adds an additional
constraint to the layout algorithm.
Instead of only avoiding overlaps, now we have the
constraint of placing the words in the same position across the clouds.
In order to do so, we present a layout algorithm that iteratively generates valid
word clouds changing the location of the shared
words to make them converge to the same position in all clouds.
We will refer to this procedure as the iterative layout algorithm, 
which is shown in Algorithm~\ref{alg:iterative}.

In particular, the iterative layout algorithm works by repeatedly calling the spiral algorithm (Section~\ref{sec:single}) with different desired locations 
for the shared words. At the first iteration, 
the desired locations are set randomly, in the same way we did for a single cloud.
Subsequently, the new desired locations are chosen by averaging 
the previous final locations of the word in the different clouds.
That is, the new desired location for word $w$ is
$p'_{w} = {|\mathcal{V}_w|}^{-1} \sum_{v_j \in \mathcal{V}_w} p_{wj}.$
Thus, the new desired locations are the same for all clouds $v_j \in \mathcal{V}_w$, $p'_{wj} = p'_{w}$. 
Changing the locations of shared words might introduce new overlaps, so we run
the Spiral Algorithm again to remove any overlaps.

In principle, this process would be repeated until the final locations are the same as the desired ones, that is, when the Spiral Algorithm does not modify the given positions. At that point all shared words will be in precisely identical positions
across the clouds.
However, this process does not always converge, so in practice, we stop after a fixed number of iterations.

However, in practice we find a serious problem with the iterative algorithm.
The algorithm tends to move words far away 
from the center, because this makes it easier to place shared words in the same position across clouds. 
This results in sparse layouts with excessive white space that are visually
unappealing.

\begin{algorithm}[tb]
\caption{Iterative Layout Algorithm}\label{alg:2}\label{alg:iterative}
\begin{algorithmic}[1]
\Require Storm $v_i = (W_i, \{ c_{iw} \}, \{ s_{iw} \})$ without positions
\Ensure Word storm $\{v_1,\dots,v_N\}$ with positions
\For{$i \in \{1,\dots,N\}$}
	\State $\bp_i \gets$ \Call{SpiralAlgorithm}{$W_i$}
\EndFor
\While{Not Converged $\wedge$ count $<$ Max Iteration }
	\For{$i \in \{1,\dots,N\}$}
	\State $p'_{iw} \gets \frac{1}{|\mathcal{V}_w|}\sum_{v_j \in \mathcal{V}_w} p_{jw}, \quad \forall w \in W_i$
		\State $\bp_i \gets$ \Call{SpiralAlgorithm}{$W_i$, $\bp'_i$}
	\EndFor
	\State count = count + 1
\EndWhile
\end{algorithmic}
\end{algorithm}

\subsection{Coordinated Layout: Gradient Approach}
\label{sec:opt}
In this section,
 we present a new method to build a storm by solving an optimization problem.
This will provide us with additional flexibility to incorporate
aesthetic constraints into storm construction, because we can incorporate
them as additional terms in the objective function.
This will allow us to avoid the unsightly sparse layouts which
are sometimes produced by the iterative algorithm.

We call the objective function the Discrepancy Between Similarities (DBS).
The DBS is a function of the set of clouds $\{ v_1, \dots, v_N \}$
and the set of documents $\{ u_1, \dots, u_N \}$,
and measures how well the storm fits the document corpus.
It is:
\begin{equation}
\begin{aligned}
f_{ u_1, \dots, u_N }(v_1,\dots,v_N)= & \sum_{1\leq i< j\leq N} (d_u(u_i,u_j) - d_v(v_i,v_j))^2  \\
& + \sum_{1\leq i\leq N} c(u_i,v_i), %
\end{aligned} \label{dbs}
\end{equation}
where $d_u$ is a distance metric between documents and $d_v$ a metric
between clouds.
The DBS is to be minimized as a function of $\{ v_i \}$.
The first summand, which we call stress, formalizes the idea that similar documents should have similar clouds and different documents, different clouds.
The second summand uses a function that we call the \emph{correspondence function}
 $c(\cdot,\cdot)$, which should be chosen to ensure that each cloud $v_i$ is a good representation of its document $u_i$.

The stress part of the objective function is inspired by multidimensional scaling (MDS).
MDS is a method for dimensionality reduction of high-dimensional data \citep{mdsborg}.
Our use of the stress function is slightly different than is common,
because instead of projecting the documents onto a low-dimensional space,
such as $\R^2$,
 we are mapping documents to the space of word clouds.
 The space of word clouds is itself high-dimesionsal, and  indeed, might 
 have greater dimension than the original space. 
Additionally, the space of word clouds is not Euclidean because of 
the non-overlapping constraints. 

For the metric $d_u$ among documents, we use Euclidean distance.
For the dissimilarity function $d_v$ between clouds, we use
$$
d_v(v_i,v_j) = \sum_{w\in \mathcal{W}}(s_{iw}-s_{jw})^2 + \kappa\sum_{w \in W_i\cap
W_j}(x_{iw}-x_{jw})^2+(y_{iw}-y_{jw})^2,
$$
where $\kappa \geq 0$ is a parameter that determines the strength of each part. 
Note that the first summand considers all words in either cloud,
and the second only the words that appear in both clouds.
(If a word does not appear in a cloud, we treat its size as zero.)
The intuition is that clouds are similar if their words have similar sizes
and locations.
Also note that, in contrast to the previous layout algorithm, by optimizing this function we will also determine the words' sizes.

The difference between the objective functions for MDS and DBS
is that
the DBS adds the correspondence function $c(u_i,v_i)$.
In MDS, the position of a data point in the target space is not interpretable on its own, but only relative to the other points.
In contrast, in our case each word cloud must accurately represent its document.
Ensuring this is the role of the correspondence function. In this work we use\begin{equation}
c(u_i,v_i) = \sum_{w\in W_i}(u_{iw} - s_{iw})^2,  \label{correspondence}
\end{equation}
where recall that  $u_{iw}$ is the $tf$ of word $w$.

We also need to add additional terms to ensure that words do not overlap, and to 
favor compact configurations.
We introduce these constraints as two penalty terms.
When two words overlap, we add a penalty proportional to the square
of the the minimum distance required to separate
them; call this distance $O_{i;w,w'}$.
We favor compactness by adding a penalty proportional to the 
the squared distance from each word towards the center; by convention
we define the origin as the center, so this is simply the norm of
word's position.

Therefore, the final objective function that we use to lay out word storms 
in the gradient based method is
\begin{multline}
g_\lambda(v_1,\dots,v_N) =  f_{ u_1, \dots, u_N }(v_1,\dots,v_N) + \\
 \lambda \sum_{i=1}^N \sum_{w,w' \in W_i} O_{i;w,w'}^2 + \mu  \sum_{i=1}^N \sum_{w \in W_i} || p_{iw} ||^2,
 \label{dbscons}
\end{multline}
where $\lambda$ and $\mu$ are parameters that determine the strength of
the overlap and compactness penalties, respectively.

We optimze \eqref{dbscons} by solving a sequence of optimization problems for increasing values 
 $\lambda_0 < \lambda_1 < \lambda_2 < \ldots$
 of the overlap penalty. We increase $\lambda$
exponentially until no words overlap in the final solution.
Each subproblem is minimized using gradient descent, initialized 
from  the solution of the previous subproblem.

\subsection{Coordinated Layout: Combined Algorithm}
\label{sec:comb}

The iterative and gradient algorithms have complementary strengths.
The iterative algorithm is fast, but as it does not enforce the compactness of the clouds, the words drift away from the center.
On the other hand, the gradient method is able to create compact clouds, but it requires 
many iterations to converge and the layout strongly depends on the initialization.
Therefore we combine the two methods, using the final result of the iterative
algorithm as the starting point for the gradient method. 
From this initialization, the gradient method converges much faster,
because it starts off without overlapping words.
The gradient method tends to improve the initial layout significantly, because it pulls 
words closer to the center, creating a more compact layout.
Also, the gradient method tends to pull together the locations of shared words 
for which the iterative method was not able to converge to a single position.

\begin{figure*}[p]
\centering
\subfloat[Electronic Materials]{\fbox{\includegraphics[width=0.31\textwidth]{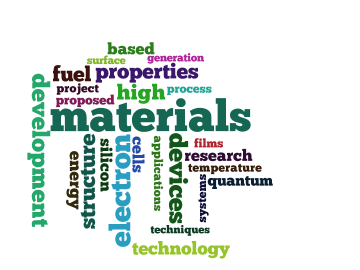}}}\hspace{1pt}
\subfloat[Metals and Alloys]{\fbox{\includegraphics[width=0.31\textwidth]{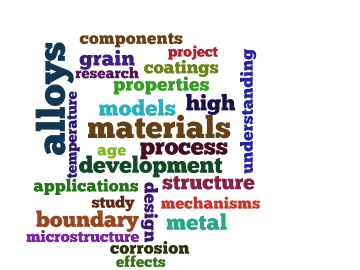}}}\hspace{1pt}
\subfloat[Photonic Materials]{\fbox{\includegraphics[width=0.31\textwidth]{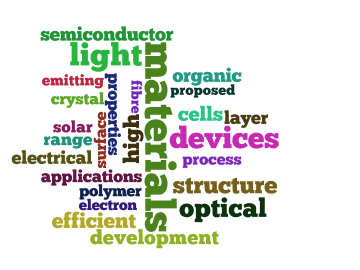}}}\vspace{1pt}
\subfloat[Structural Ceramics and Inorganics]{\fbox{\includegraphics[width=0.31\textwidth]{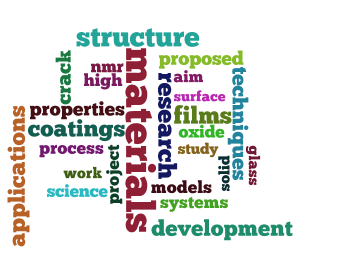}}}\hspace{1pt}
\subfloat[Structural Polymers and Composites]{\fbox{\includegraphics[width=0.31\textwidth]{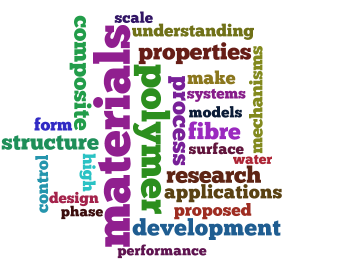}}}\hspace{1pt}
\subfloat[Mathematical Sciences]{\fbox{\includegraphics[width=0.31\textwidth]{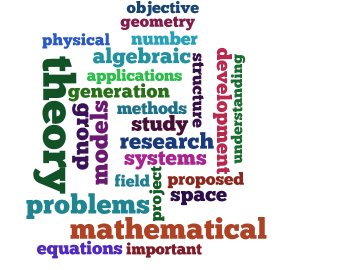}}}
\caption{Independent Clouds visualizing six EPSRC Scientific Programmes.
These programmes are also represented in Figure~ \ref{fig:mostdiffcoor}}
\label{fig:mostdiffindp}
\end{figure*}

\begin{figure*}[p]
\centering
\subfloat[Electronic Materials]{\fbox{\includegraphics[width=0.31\textwidth]{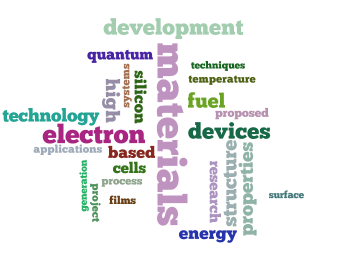}}}\hspace{1pt}
\subfloat[Metals and Alloys]{\fbox{\includegraphics[width=0.31\textwidth]{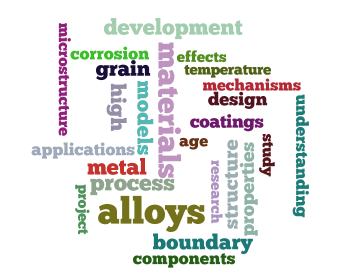}}}\hspace{1pt}
\subfloat[Photonic Materials]{\fbox{\includegraphics[width=0.31\textwidth]{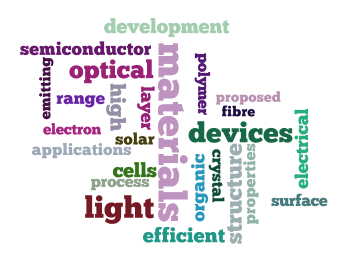}}}\vspace{1pt}
\subfloat[Structural Ceramics and Inorganics]{\fbox{\includegraphics[width=0.31\textwidth]{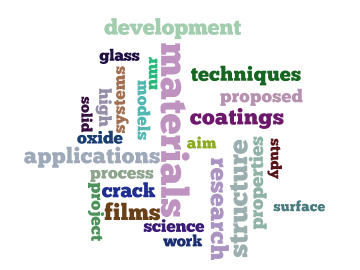}}}\hspace{1pt}
\subfloat[Structural Polymers and Composites]{\fbox{\includegraphics[width=0.31\textwidth]{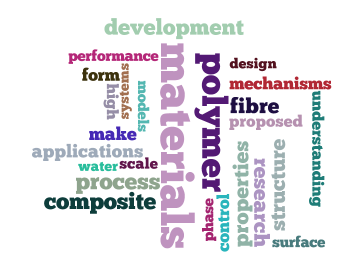}}}\hspace{1pt}
\subfloat[Mathematical Sciences]{\fbox{\includegraphics[width=0.31\textwidth]{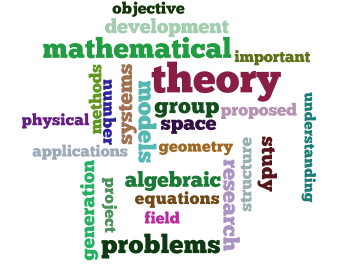}}}
\caption{Coordinated storm visualizing six EPSRC Scientific Programmes.
These programmes are also represented as independent clouds in Figure~\ref{fig:mostdiffindp}. Compared to that figure, it is much
easier to see the differences between clouds.}
\label{fig:mostdiffcoor}
\end{figure*}

\begin{table*}[tb]
\centering
\begin{tabular}{l*{1}{r}{c}{c}}
\centering
	    &  Time (s) & Compactness (\%)  & Accuracy (\%) \\
\hline
Lower Bound			& - & -  		& 26.5 \\
Independent Clouds   		& 143.3   & 35.12	& 23.4 \\
Coordinated Storm (Iterative)   & 250.9   & 20.39  	& 54.7 \\
Coordinated Storm (Combined)   	& 2658.5  & 33.71   	& 54.2 \\ 
Upper Bound			& - & -	 		& 67.9 \\
\end{tabular}
\caption{Comparison of the results given by different algorithms using the automatic evaluation.  \label{tab:auteval}}
\end{table*}

\section{Evaluation}
The evaluation is divided in three parts: a qualitatively analysis, an automatic analysis and a user study.
We use two different data sets.
We used the scientific papers presented in the ICML 2012 conference,
 where we deployed a storm on the main conference Web site
to compare the presented papers and help the people decide among sessions\footnote{\url{http://icml.cc/2012/whatson/}}.

Second, we also use a data set provided by the Research Perspectives project\footnote{Also see \url{http://www.researchperspectives.org}} \cite{khalifa13multiobjective},
a project that aims to offer a visualization of the research portfolios
of funding agencies.
This data set contains the abstracts of the proposals for funded research grants
from various funding agencies.
We use a corpus of 2358 abstracts from the UK's Engineering and Physical
Sciences Research Council (EPSRC).
Each grant belongs to exactly one of the following programmes: 
Information and Communications Technology (626 grants),
Physical Sciences (533),
Mathematical Sciences (331),
Engineering (317),
User-Led Research (291) and
Materials, Mechanical and Medical Engineering (264).
Each of these top-level programmes as several subprogrammes
that correspond to more specific research areas.

\subsection{Qualitative Analysis}
In this section, we discuss the presented storms qualitatively,
focusing on the additional information that is 
apparent from coordinated storms compared to independently built clouds.

First, we consider a storm that displays six research programmes from EPSRC programmes, 
five of which are different subprogrammes of material sciences and the sixth one
is the mathematical sciences programme.
For this data set we present both
a set of independent clouds (Figure~\ref{fig:mostdiffindp}) 
and a storm generated by the combined algorithm (Figure~\ref{fig:mostdiffcoor}).
From either set of clouds, we can get superficial idea of the corpus.
We can see the most important words such as ``materials'', which appears in the first five clouds, 
and some other words like ``alloys'', ``polymer'' and ``mathematical''. 
However, it is hard to get more information than this from the independent clouds.

On the other hand, by looking at the coordinated storm we can detect more properties of the corpus.
First, it is instantly clear that the first five documents are similar and that the sixth one is the most different from all the others. This is because the storm reveals the shared structure in the documents, formed by shared words such as ``materials'', ``properties'' and ``applications''. 
Second, we can easily tell the presence or absence of words across clouds
because of the consistent attributes and locations.
For example, we can quickly see that ``properties'' does not appear in the sixth cloud or that ``coatings'' only occurs in two of the six.
Finally, the transparency of the words allows us to spot the informative terms 
quickly, such as ``electron'' (a), ``metal'' (b), ``light'' (c), ``crack''(d), ``composite''(e) and ``problems''(f).
All of these term are informative of the document content but are
difficult to spot in the independent clouds of Figure~\ref{fig:mostdiffindp}.
Overall, the coordinated storm seems to offer a more rich and comfortable representation that allows deeper analysis than the independently
generated clouds.

Similarly, from the ICML 2012 data set, Figure~\ref{fig:icml} shows
a storm containing all the papers from a single conference session.
It is immediately apparent from the clouds that the session discusses optimization algorithms.
It is also clear that the papers (c) and (d) are very related since they share a lot of words such as ``sgd'', ``stochastic'' and ``convex'' which results in a similar layouts. 
The fact that shared words take similar positions can also force unique words into
similar positions as well, which can 
make it easy to find terms that differentiate the clouds.
For example, we can see how ``herding'' (f), ``coordinated'' (g) and ``similarity'' (h) 
are in the same location or ``semidefinite'' (a), ``quasi-newton'' (b) and ``nonsmooth'' (d) are in the same location.

Finally, Figures \ref{fig:epsrc1} and \ref{fig:epsrc2} show
an example of a hierarchical set of storms generated from the EPSRC
grant abstracts.
Figure \ref{fig:epsrc1} presents a storm created by grouping all abstracts by
their top level scientific program.
There we can see two pairs of similar programmes: Chemistry and Physical Sciences; and Engineering and Information Communication and Technology.
In Figure \ref{fig:epsrc2}, we show a second storm composed a six individual
grants from the Complexity programme (Cloud (e) in Figures~\ref{fig:epsrc1}).
It is interesting to see how big words in the top level such as ``complex'', ``systems'', ``network'' and ``models'' 
appear with different weights in the grant level.
In particular, the term ``complex'', that it is rare when looking at the top level, appears everywhere inside the complexity programme.
Because of our use of transparency, this term is therefore prominent in the top
level storm but less noticeable in the lower level storm.

\subsection{Automatic Evaluation}
Apart from evaluating the resulting storm qualitatively, we propose a method to evaluate word storm algorithms automatically. 
The objective is to assess how well the relations among documents are represented in the clouds.
The motivation is similar in spirit to the celebrated
BLEU measure in machine translation \cite{bleu}:
By evaluating layout algorithms 
with an automatic process rather than conducting a user study, the process can be faster and inexpensive, allowing rapid comparison of algorithms.

Our automatic evaluation requires a corpus of labelled documents, e.g.,
with a class label that indicates their topics.
The main idea is: If the visualization is faithful to the documents,
then it should be possible to classify the documents using the pixels
in the visualization rather than the words in the documents.
So we use classification accuracy as a proxy measure
for visualization fidelity.

In the context of word storms,
 the automatic evaluation consists of: (a) generating a storm from a labelled corpus 
 with one cloud per cloud, (b) training a document 
 classifier using the pixels of the clouds as attributes and (c) testing the classifier on a held out set
to obtain the classification accuracy.  More faithful visualizations are expected to have better
classification accuracy.
We use the Research Perspectives EPSRC data set with the research
programme as class label.
Thus, we have a single-label classification problem with 6 classes.
The data was randomly split into a training and test set using an 80/20 split.
We use the word storm algorithms to create one cloud per abstract,
so there are  2358 clouds in total.
We compare three layout algorithms: 
(a) creating the clouds independently using the Spiral Algorithm, 
which is our baseline; 
(b) the iterative algorithm with 5 iterations and 
(c) the combined algorithm, using 5 iterations
of the iterative algorithm to initialize the gradient method.

We represent each cloud by a vector of the RGB values of its pixels.
To reduce the size of this representation, we perform feature selection, 
discarding features with zero information gain.  
We classify the clouds by using support vector machines with normalized polynomial kernel\footnote{The classification is performed by using the SMO implementation of Weka}.  

In order to put the classification accuracy into context, we present a 
lower bound obtain if all instances are
classified as the largest class (ICT), which produces an accuracy of 26.5\%.
To obtain an upper bound,  we classifying the documents directly using bag-of-words features from the text, which should perform
better than transforming the text into a visualization.
Using a support vector machine, this yields an accuracy of 67.9\%.

Apart from the classification accuracy, we also report the running time of the layout
algorithm (in seconds)\footnote{All experiments were run on a 3.1 GHz Intel Core i5 server with 8GB of RAM.}, and, as a simple aesthetic measure,
the compactness of the word clouds. 
We compute the compactness by taking the minimum bounding box of the cloud and calculating the percentage of non-background pixels. 
We use this measure because informally we noticed 
that more compact clouds tend to be more visually appealing.

The results are shown in Table~\ref{tab:auteval}.
Creating the clouds independently is faster than any coordinated
algorithm and also  produces very compact clouds. 
However, for classification, this method is no better than random.
The algorithms to create coordinated clouds, the iterative and the combined algorithm, achieve a 54\% classification accuracy, which is significantly higher than the lower bound. 
This confirms the intuition that by coordinating the clouds, the relations among documents are better represented.

The differences between the coordinated methods can be seen in the running time and in the compactness.
Although the iterative algorithm achieves much better classification
accuracy than the baseline, this is at the cost of producing much less compact
clouds.
The combined algorithm, on the other hand, is able to match the compactness of independently built clouds (33.71\% combined and 35.12\% independent) 
and the classification accuracy of the iterative algorithm.
The combined algorithm is significantly more expensive in computation time,
although it should be noted that even the combined algorithm uses only 1.1s 
for each of the 2358 clouds in the storm.
Therefore, although the combined algorithm requires more time, it seems the best option, because the resulting storm offers good classification accuracy without losing compactness. 

A potential pitfall with automatic evaluations is that it is possible 
for algorithms to game the system, producing visualizations that score
better but look worse.
This has arguably happened in machine translation, in which BLEU has been
implicitly optimized, and possibly 
overfit, by the research community for many years.
For this reason it is important to combine 
Furthermore, in our case, 
none of our the algorithms optimize the classification accuracy directly
but instead follow very different considerations.
But the concern of ``research community overfitting'' is one to take
seriously if automated evaluation of visualization is more widely
adopted.

\begin{table*}[tb]
\centering
\begin{tabular}{l l c c}
&  & Independent clouds & Coordinated Storm \\ \hline
\multirow{3}{5cm}{Select clouds with word ``technology''} 
 & Precision (\%)	& 90 & 100 \\
 & Recall (\%)		& 65 & 85\\
 & Time (s)		& 51 $\pm$ 23 & 36 $\pm$ 10  \\ %
\midrule
\multirow{3}{5cm}{Select clouds without word ``energy''} 
 & Precision (\%)	& 90 & 93\\
 & Recall (\%)		& 85 & 95\\
 & Time (s)		& 56 $\pm$ 18 & 40 $\pm$ 14  \\ %
\midrule
\multirow{3}{5cm}{Select clouds with words ``models'', ``network'' and ``system''} 
 & Precision (\%)	& 75 & 90 \\
 & Recall (\%)		& 90 & 100\\
 & Time (s)		& 87 $\pm$ 35 & 124 $\pm$ 46  \\ %
\midrule
\multirow{2}{5cm}{Select the most different cloud} 
 & Accuracy (\%)	& 30 & 90 \\
 & Time (s)		& 36  $\pm$ 12 & 23 $\pm$ 10  \\ %
\midrule
\multirow{2}{5cm}{Select the most similar pair clouds} 
 & Accuracy (\%)	& 10 & 70 \\
 & Time (s)		& 54  $\pm$ 23 & 75 $\pm$ 19 
\end{tabular}
\caption{Results of the user study. 
Accuracy on the last two questions, which required comparing documents, 
is much higher for the users that were presented with a coordinated storm.}
\label{tab:study}
\end{table*}

\subsection{User Study}

In order to confirm our results using the automatic evaluation, we conducted a pilot user study comparing the standard independent word clouds with coordinated storms created by the combined algorithm. 
The study consisted of 5 multiple choice questions. In each of them, the users were presented with six clouds and were asked to perform a simple task. 
The tasks were of two kinds: checking the presence of words and comparing documents.
The clouds for each question were generated either as independent clouds or a coordinated storm. 
In every question, the user received one of the two versions randomly\footnote{The random process ensured that we would have the same number of answers for each method}. 
Although users were told in the beginning that word clouds had been built using different methods, the number of different methods was not revealed, the characteristics of the methods were not explained and they did not know which method was used for each question. 
Moreover, in order to reduce the effect of possible bias factors, the tasks were presented in a random order and the 6 clouds in each question were also sorted randomly.
The study was taken by 20 people, so each question was answered 10 times using the independent clouds and 10 times using a coordinated storm.

Table \ref{tab:study} presents the results of the study. 
The first three questions asked the users to select the clouds that contained or lacked certain words.
The results show that although the precision and recall are high in both cases and the differences are small, 
the coordinated storm always has a higher score than the independent clouds.
This might be because the structured layout helped the users to find words, 
even though the users did not know how the storms were laid out.

The last two questions asked the users to compare the documents and to select ``the cloud that is most different from all the others''
 and ``the most similar pair of clouds''. 
In the first case, 
two clouds had  a cosine similarity\footnote{The documents were taken using the bag of words representation with frequencies. The cosine similarity was computed twice: considering all words and only considering the top 25 words included in the cloud.} lower than 0.3 with all the others, while all others pair had a similarity higher than 0.5.
In the last question, the most similar pair of clouds had a cosine similarity of 0.71, 
while the score of the second most similar was 0.48.
As these questions only have a correct answer, the measure used is the accuracy, instead of the precision and recall.

The results for the last two questions show that the coordinated storm outperforms the independent clouds. 
While a 90\% and a 70\% of the users presented with the coordinated version answered correctly, only a 30\% and a 10\% did so with the independent version. 
This confirms that coordinated storms allow the users to contrast the clouds and understand their relations,
while independent clouds are misleading in these tasks.

Although the sample size is small, results favour the coordinated storm. 
In particular, when the users are asked to compare clouds, the differences in
user accuracy are extremely large.
Regarding the answering time, the
differences between the two conditions are not significant.

\section{Related Work}

Word clouds were inspired by tag clouds, 
which first appeared as an attractive way to summarize and browse a user-specified folksonomy.
Originally, the tags were organized in horizontal lines and sorted by alphabetical order, 
a layout that is still used in many websites such as Flickr and Delicious.
Word clouds extend this idea to document visualization.
Of the many word cloud generators, one of the most popular is Wordle 
\citep{feinberg10beautiful, partWordle}, 
which produces particularly appealing layouts.

However, in contrast to visualizing a single document, the topic of visualizing corpora has received much less attention.
Several research has proposed to create the clouds by using different importance measures, such as the $tf*idf$ \citep{impTag} or the relative frequency when only the relations of a single document have to be analysed \citep{beatvis, neoformix}. 
Nevertheless, without a different layout algorithm clouds are still difficult to compare because they do not attempt to follow the coordination of similarity principle and shared words are hard to find.

\citet{parallel} presented Parallel Tag Clouds, 
a method that aims to make comparisons easier by representing the documents as lists.
The closest related work was presented by \citet{weiwei}, which was later improved
by \citet{semantic}.
This work proposes using a sequence of word clouds along with a trend chart 
to show the evolution of a corpus over time.
They present a new layout algorithm with the goal of keeping semantically
similar words close to each other in each cloud.
This goal is very different from that of our work:
Preserving semantic relations between words within a cloud
is different than coordinating similarities across clouds,
and does not necessarily result in similar documents being represented by similar clouds.

\section{Conclusions}

We have introduced the concept of word storms, which is a
group of word clouds designed for the visualization of a corpus
of documents.
We presented a series of principles for effective storms, arguing
that the clouds in a storm should be built in a coordinated fashion
to facilitate comparison.
We presented a novel algorithm that builds storms in a coordinated fashion,
placing shared words in a similar location across clouds,
so that similar documents will have similar storms.
Using both an automatic evaluation and a user study,
we showed that coordinated storms were markedly superior to independent
word clouds for the purposes of comparing and contrasting documents.

%

\section{Acknowledgments}

We thank Mike Chantler and Fraser Halley for kindly providing
access to the research grant data.
This work was supported by the Engineering and Physical Sciences Research Council [grant number EP/J00104X/1].

\bibliographystyle{abbrvnat}
\bibliography{refs}  %

\begin{thebibliography}{14}
\providecommand{\natexlab}[1]{#1}
\providecommand{\url}[1]{\texttt{#1}}
\expandafter\ifx\csname urlstyle\endcsname\relax
  \providecommand{\doi}[1]{doi: #1}\else
  \providecommand{\doi}{doi: \begingroup \urlstyle{rm}\Url}\fi

\bibitem[Borg and Groenen(2005)]{mdsborg}
I.~Borg and P.~Groenen.
\newblock \emph{{Modern Multidimensional Scaling: Theory and Applications}}.
\newblock Springer, 2005.

\bibitem[Chernoff(1973)]{chernoff73faces}
H.~Chernoff.
\newblock The use of faces to represent points in k-dimensional space
  graphically.
\newblock \emph{Journal of the American Statistical Association}, 68\penalty0
  (342):\penalty0 361--368, 1973.

\bibitem[Clark(2008)]{neoformix}
J.~Clark.
\newblock Clustered word clouds - neoformix, april 2008.
\newblock URL \url{http://www.neoformix.com/}.

\bibitem[Collins et~al.(2009)Collins, Viegas, and Wattenberg]{parallel}
C.~Collins, F.~B. Viegas, and M.~Wattenberg.
\newblock Parallel tag clouds to explore and analyze faceted text corpora.
\newblock pages 91--98, Oct. 2009.
\newblock \doi{10.1109/VAST.2009.5333443}.
\newblock URL \url{http://dx.doi.org/10.1109/VAST.2009.5333443}.

\bibitem[Cui et~al.(2010)Cui, Wu, Liu, Wei, Zhou, and Qu]{weiwei}
W.~Cui, Y.~Wu, S.~Liu, F.~Wei, M.~Zhou, and H.~Qu.
\newblock Context-preserving, dynamic word cloud visualization.
\newblock \emph{IEEE Computer Graphics and Applications}, 30:\penalty0 42--53,
  2010.
\newblock ISSN 0272-1716.
\newblock \doi{http://doi.ieeecomputersociety.org/10.1109/MCG.2010.102}.

\bibitem[Feinberg(2010)]{feinberg10beautiful}
J.~Feinberg.
\newblock Wordle.
\newblock In J.~Steele and N.~Iliinsky, editors, \emph{Beautiful Visualization
  Looking at Data through the Eyes of Experts}, chapter~3. O'Reilly Media,
  2010.

\bibitem[Hassan-Montero and Herrero-Solana(2006)]{impTag}
Y.~Hassan-Montero and V.~Herrero-Solana.
\newblock Improving tag-clouds as visual information retrieval interfaces.
\newblock In \emph{InScit2006: International Conference on Multidisciplinary
  Information Sciences and Technologies}, 2006.
\newblock URL \url{http://nosolousabilidad.com/hassan/improving_tagclouds.pdf}.

\bibitem[Khalifa et~al.(2013)Khalifa, Corne, Chantler, and
  Halley]{khalifa13multiobjective}
O.~Khalifa, D.~Corne, M.~Chantler, and F.~Halley.
\newblock Multi-objective topic modelling.
\newblock In F.~F. Purshouse and S.~Greco, editors, \emph{Evolutionary
  Multi-Criterion Optimization (EMO 2013)}, 2013.

\bibitem[Papineni et~al.(2002)Papineni, Roukos, Ward, and Zhu]{bleu}
K.~Papineni, S.~Roukos, T.~Ward, and W.-J. Zhu.
\newblock Bleu: a method for automatic evaluation of machine translation.
\newblock In \emph{Proceedings of the 40th Annual Meeting on Association for
  Computational Linguistics}, ACL '02, pages 311--318, Stroudsburg, PA, USA,
  2002. Association for Computational Linguistics.
\newblock \doi{10.3115/1073083.1073135}.
\newblock URL \url{http://dx.doi.org/10.3115/1073083.1073135}.

\bibitem[Steele and Iliinsky(2010)]{beatvis}
J.~Steele and N.~Iliinsky.
\newblock \emph{{Beautiful Visualization: Looking at Data through the Eyes of
  Experts}}.
\newblock Oreilly \& Associates Inc, 2010.
\newblock ISBN 1449379869.

\bibitem[Tufte(1997)]{Tufte1997}
E.~R. Tufte.
\newblock \emph{Visual Explanations: Images and Quantities, Evidence and
  Narrative}.
\newblock Graphics Press LLC, 1997.

\bibitem[Tufte(2001)]{Tufte2001}
E.~R. Tufte.
\newblock \emph{The Visual Display of Quantitative Information}.
\newblock Graphics Press LLC, 2nd edition, 2001.

\bibitem[Viegas et~al.(2009)Viegas, Wattenberg, and Feinberg]{partWordle}
F.~B. Viegas, M.~Wattenberg, and J.~Feinberg.
\newblock Participatory visualization with wordle.
\newblock \emph{IEEE Transactions on Visualization and Computer Graphics},
  15\penalty0 (6):\penalty0 1137--1144, Nov. 2009.
\newblock ISSN 1077-2626.
\newblock \doi{10.1109/TVCG.2009.171}.
\newblock URL \url{http://dx.doi.org/10.1109/TVCG.2009.171}.

\bibitem[Wu et~al.(2011)Wu, Provan, Wei, Liu, and Ma]{semantic}
Y.~Wu, T.~Provan, F.~Wei, S.~Liu, and K.-L. Ma.
\newblock Semantic-preserving word clouds by seam carving.
\newblock \emph{Comput. Graph. Forum}, 30\penalty0 (3):\penalty0 741--750,
  2011.

\end{thebibliography}
\end{document}